# Data Bits in Karnaugh Map and Increasing Map Capability in Error Correcting

Pouya Pezeshkpour, and Mahmoud Tabandeh
*Electrical Engineering Department, Sharif University of Technology, Tehran, Iran.*
emails: pezeshkpour_pouya@ee.sharif.edu, tabandeh@sharif.edu

*Abstract*---To provide reliable communication in data transmission, ability of correcting errors is of prime importance. This paper intends to suggest an easy algorithm to detect and correct errors in transmission codes using the well-known Karnaugh map. Referring to past research done and proving new theorems and also using a suggested simple technique taking advantage of the easy concept of Karnaugh map, we offer an algorithm to reduce the number of occupied squares in the map and therefore, reduce substantially the execution time for placing data bits in Karnaugh map. Based on earlier papers, we first propose an algorithm for correction of two simultaneous errors in a code. Then, defining specifications for empty squares of the map, we limit the choices for selection of new squares. In addition, burst errors in sending codes is discussed, and systematically code words for correcting them will be made.

## I. INTRODUCTION

Detection and correction of bit errors in transmission of code has been investigated using Hamming code [1]. Hamming code is capable of correcting one or detecting 2 errors. Using similarities of the concept of this technique and the Karnaugh map philosophy, efforts were made to detect and correct multiple errors in a code, and an effective method was provided for correction of two bit errors in a code [1]. In this paper, using a Karnaugh map, placement and arrangement of 4 data and 7 parity bits to produce a code correcting single and double errors were studied. This procedure was extended successfully in [2] adding another feature of correcting 3 adjacent errors. A weakness of that technique was that obtaining the desired code required a large amount of time. Therefore, limiting selection range of the candidate squares in the Karnaugh map would result in substantial reduction in computer time.

Similar approaches for error detecting and correcting codes was conducted in [3], [4] and [5] but they did not use the Karnaugh map for this purpose.

The object of this research was to find an algorithm to provide boundaries for computer selection range and to evaluate unused squares of the Karnaugh map for eventually placing information bits by studying them more closely and finally to create a Karnaugh map with more capabilities .

Let's start by reviewing some facts discussed in references [1] and [2]. An n variable Karnaugh map is subdivided into $2^n$ squares and since both its rows and columns are arranged using Gray codes, two adjacent squares in the map differ only by one variable in their K-code, a binary representation for the square. That is, the binary code representing every adjacent square to a specific square will have a Hamming distance of one with that square. To detect and correct errors in a received code, we divide the set of all code bits into different subsets represented by $s_k$ ($k\epsilon\{1,...,7\}$) and consider one parity ($P_k$) and two or more data bits as their members. The value of the parity bit in each subset is determined in such a way that the sum (modulo 2) of all members of that subset will be equal to zero (case of even parity). At the receiving end, by comparing the value of parity bit in each subset with the expected value (even or odd), we can figure out which subset contains an error and then specify the corresponding check bit in the Karnaugh map. The code produced in this way is called the K-code for the specific error. It also specifies a square in the Karnaugh map associated with that error. In this way, one square of the Karnaugh map is assigned to each error we wish to correct. For correcting multiple errors, we have to assign one square to every error we plan to correct.

In the following, in section 2 we start making some definitions. Then, a technique for placing data bits in the Karnaugh map will be given in section 3. Investigation on empty squares of the map for potential selection will be done in section 4. Finally, concluding remarks will be given in section 5.

## II. DEFINITIONS

In the following definitions, we refer to reference [1] and [2] and use the same definitions which provided in them.
1- The code specifying one square in a Karnaugh map is referred to as the K-code for that square.
2- Assume a data bit $X_i$, $i\epsilon\{1,...,4\}$ represented by a K-code, we define "First order side squares" as the squares with the Hamming distance of 1 from $X_i$. Note that these adjacent squares correspond to $X_iP_j$ ($j\epsilon\{1,...,7\}$) which are actually the images of $X_i$ against $P_j$ axes.

| $S_4$ $S_2$ $S_3$ $S_1$ | 00 | 01 | 11 | 10 |
|---|---|---|---|---|
| 00 |  | 2 | 1 | 2 |
| 01 | 2 | 1 | X | 1 |
| 11 |  | 2 | 1 | 2 |
| 10 |  |  | 2 |  |

Fig. 1. Displaying First and Second order side squares of a given $X$ in a Karnaugh map.

3- In the same way, "Second order side squares" are those squares with a Hamming distances of 2 from any $X_i$ placed in the Karnaugh map.
4- A square whose K-code consists of $m$ non zero elements and, of course, 7-$m$ zeros ($m \epsilon \{1, ..., 7\}$), is referred to as $N_m$.
5- The side squares (either First or Second order) from two different $X$'s that are placed over each other are referred to as double weight squares.
6- The procedure in which $X_1$ is placed in $N_i$ and $X_2$ is placed in $N_j$ with a Hamming distance of $p$ from each other and $X_3$ is placed in $N_k$ with a Hamming distance of $q$ from $X_1$ and Hamming distance of $r$ from $X_2$ is referred to as a $S_{ijk}^{pqr}$ ($S_i$ will be used before placing $X_2$ and $X_3$ in the map and $S_{ij}^p$ will be used before placing $X_3$ in the map).

The Karnaugh map of the extended Hamming code with the four parity check bits is shown in Fig. 1. As can be seen in this figure, $X$ is placed in $N_3$ with $S_4S_3S_2S_1 = 1011$. The squares labeled as "1" are the First order side squares of $X$. That is, the squares having a distance of 1 from $X$. The squares labeled as "2" are the Second order side squares of $X$.

## III. A TECHNIQUE FOR PLACING DATA BITS FOR DOUBLE-ERROR CORRECTING MAP

As mentioned before, to place data bits in desired squares, we need to find an algorithm to reduce the amount of computation time. To do so, we have to find a unique property that prioritizes some squares over others for being selected as data bit placement. Our aim is to place 4 data and 7 parity bits and their side squares in the map, in order to cover all the possible situations of 1 and 2 bit errors in an eleven bit code in which each error case would be assigned to only one square. To achieve this, let's assume that data bits are represented by their K-code $X_i$, and parity bits by $P_k$ ($k \in \{1, ..., 7\}, i, j \in \{1, ..., 4\}$). Of course, $X_j$ cannot be placed in an $X_i$ or $P_k$ square. Neither can they be placed on their First or Second order side squares.

Suppose $X_1$, $X_2$ and $X_3$ are placed on the map. Suppose also that we can find a square not coinciding with a first or second order side squares of any of the already existing $X_i$'s and $P_k$'s. Then, placing $X_4$ in such a square makes it possible that all the $X_4P_k$ squares can be determined in the map without any overlap. Now our aim is to keep unused high priority squares after $X_1$, $X_2$ and $X_3$ have been placed in the map. To keep the largest number of free squares (not occupied by First or Second order side squares), we try to increase the occupancy of squares already taken by First or Second order side squares. That is, we try to choose $X_i$'s in such a manner to overlap as many as possible new First and Second order side squares with the existing ones. Given two data bits $X_i$ and $X_j$ placed on the map, for all $i, j \epsilon \{1, ..., 4\}$ First order side squares of $X_i$ and $X_j$ are not allowed to be placed over each other. Now, in order to have the statistics, we will count the side squares for $X_i$ ($i \epsilon \{1, ..., 4\}$). Obviously, the more double weight squares we could place in a certain square, the higher priority it would have. We know that, by definition all $P_k$'s are placed in $N_1$. Also, $P_kP_m$ are all placed in $N_2$ ($k, m \epsilon \{1, ..., 7\} \& k \neq m$) and Second order side squares of $P_k$ are placed in $N_3$ and $N_1$. Therefore, $X_i$ has to be placed in $N_p$ where $p \geq 4$. Now, we start placing $X_i$'s in $N_4$, and count relevant double weight side squares. In $S_4$ (see definitions), $X_1$'s First order side squares will be located in $N_3$ and $N_5$, not coinciding with already occupied squares. In the same way, Second order side squares would be located in $N_2$, $N_4$ and $N_6$. Since no square greater than $N_3$ has been the double weight side square (whether First or Second order) of $X_1$. Double weight side squares occur only when First order side squares of $X_1$ are placed in $N_3$ where their number is equal to $\binom{4}{1} = 4$, and when Second order side squares of $X_1$ are placed in $N_2$ where their number is equal to $\binom{4}{2} = 6$. This means, placing $X_1$ in $N_4$ would result in a total of 10 double weight side squares. In $S_5$, the double weight side squares occur only when Second order side squares of $X_1$ are placed in $N_3$ their number would be equal to $\binom{5}{2} = 10$. Also, placing $X_1$ in an $N_5$ would result in a total of 10 double weight side squares which results the same as the previous situation. In $S_6$ and $S_7$, there would be no double weight squares. Therefore, since we are trying to compact errors in squares, we place $X_1$ in $N_4$ and test free squares for $X_2$ to find out about relevant side squares generated. This process will be repeated with placing $X_1$ in $N_5$. If $X_1$ and $X_2$ are placed in $N_4$, since $X_2$ cannot be placed in any of side squares of $X_1$. Here as before, we would get the number 10 for double weight side squares. Now we check to see which one of $X_2$ side squares will be placed over side squares of $X_1$ which will be added to the 10 double weight side squares we already have. In this case, either their first order side squares will double weight each other (which is not possible) or their Second order side squares will double weight each other which means that the Hamming distance of $X_1$ and $X_2$ must be equal to four. In the following, we first state and prove three theorems whose results will be useful to us to continue our procedure.

*Theorem 1:* If two squares considered for placement of $X_i$ and $X_j$ have a Hamming distance of 4, then there will be 6 Second order double weight side squares.

*Proof:* $X_i$ and $X_j$ are considered as below:

$$X_i = \sum_{l=0}^{7} a_l 2^l \quad (1)$$

$$X_j = \sum_{m=0}^{7} b_m 2^m \quad (2)$$

$a_l$ and $b_m \in \{0,1\}$

Either 4 or 5 of $a_l$'s and $b_m$'s will be equal to 1. In order to get to $X_j$ with complementing $X_i$'s values, we have to convert two of $a_l$'s that have a none zero value to 0, and two of $a_l$'s that are 0 to none zero value (Similar to $b_m$). Thus, out of these four $a_l$'s that need to be converted, each pair conversion will result in duplication of Second order side squares of $X_i$ with Second order side squares of $X_j$. Therefore, the total number of Second order double weight side squares will be equal to $\binom{4}{2} = 6$. □

*Theorem 2:* If the difference in the number of none zero elements expressing the K-codes of two squares allocated to $X_i$ and $X_j$ is equal to 1, say, if $X_i \in N_p$ and $X_j \in N_{p+1}$ and if their Hamming distance is equal to 3, then in relation to these assumptions there will be 6 double weight side squares.

*Proof:* We consider $X_i$ and $X_j$ Similar to the previous theorem. According to our assumption, either 4 of $a_l$'s and 5 of $b_m$'s or 5 of $a_l$'s and 6 of $b_m$'s are equal to 1. Since the Hamming distance between $X_i$ and $X_j$ is 3, in order to get to $X_j$ with complementing $X_i$'s values, we have to convert (with an assumption that the number of $b_m$'s that equal to 1 is larger than $a_l$'s) two of $a_l$'s that are 0 to 1, and one of $a_l$ that is 1 to 0, or convert two of $b_m$ that are 1 to 0, and one of $b_m$ that is 0 to 1.

Therefore, the total number of double weight side squares will be equal to $\binom{3}{2} + \binom{3}{1} = 6$. □

*Theorem 3:* The Hamming distance of $X_i X_j$ to either $X_i$ or $X_j$ is at least 2.

*Proof:* Since $X_i$ has a Hamming distance of at least 3 from $X_j$, there will be at least one difference in these three bits between $X_i X_j$ and either of $X_i$ or $X_j$. We also know that $X_i$ and $X_j$ are common in at least one bit whose value is equal to 1 and therefore in that bit $X_i X_j$ is equal to 0. Thus, $X_i X_j$ has at least a Hamming distance of 2 from $X_i$ and $X_j$. □

According to Theorem 1 and the fact that only one of 6 double weight side squares are located in $N_2$ (already counted), it is concluded that if $X_2$ is placed in $N_4$ in such a way to have a distance of 4 from $X_1$, the total number of double weight side squares created would be equal to 15. However, if we place $X_2$ in $N_5$, there will be at least 10 side square overlying the previously obtained side squares. In order to place $X_2$ side squares on $X_1$ side squares, either its Second order side squares have to be placed on First order side squares of $X_1$, or its First order side squares have to be placed on Second order side squares of $X_1$. In either case, the Hamming distance between $X_2$ and $X_1$ must be 3.

According to Theorem 2 and the fact that only one of the 6 common double weight side squares between $X_1$ and $X_2$ is located in $N_3$ (previously taken into account), it is concluded that if $X_2$ considered to have a distance of 3 from $X_1$ is placed in $N_5$, the total number of double weight side squares would be equal to 15. If $X_2$ is placed in $N_5$ without having a Hamming distance of 3 from $X_1$, the total number of double weight side squares will be equal to 10. However, If $X_2$ is placed in $N_6$, there will be double weight squares only if the Hamming distance from $X_1$ is 4, and in this case, according to Theorem 1, the total number of double weight squares would be equal to 6. If $X_2$ is placed in $N_7$, according to Theorem 2, the total number of double weight squares would be equal to 6. Now if we put $X_1$ in $N_5$ and $X_2$ in $N_4$, we will have 10 double weight side squares for $X_2$ and in order to place $X_1$ and $X_2$ side squares overlapping with each other, their Hamming distance should be equal to 3. In this case, according to the Theorem 1, the total number of double weight squares would be equal to 6 one of which is located in $N_3$ that was previously taken into account. Therefore, in $S_{54}^3$ the total number of double weight side squares for $X_2$ would be equal to 15. And if the Hamming distance be greater than 3, the total number of double weight side squares for $X_2$ will remain at 10. Now, if we put $X_1$ and $X_2$ in $N_5$, there will be 10 double weight squares for $X_2$ as counted before, and in order to place side squares of $X_1$ and $X_2$ over each other, their Hamming distance should be equal to 4. In this case, according to Theorem 1, the total number of double weight squares would be equal to 6 from one of which is placed in $N_3$ which was previously taken into account. Therefore, in $S_{55}^4$ the total number of double weight side squares for $X_2$ would be equal to 15. And if the Hamming distance is not 4, the total number of double weight side squares for $X_2$ will remain at 10. In $S_{56}^3$, the total number of double weight squares would be equal to 6. And if $X_1$ is placed in $N_5$, $X_2$ cannot be placed in $N_7$. Now, to get to the optimized algorithm, we either have to put both $X_1$ and $X_2$ in $N_4$ with a Hamming distance of 4, or put them in $N_5$ with a Hamming distance of 4, or put one of them in $N_5$ and the other in $N_4$ with a Hamming distance of 3. Now we have to check the placement of $X_1X_2$. Since the K-code for the square in which $X_1X_2$ should be placed is equal to XOR of the K-codes for $X_1$ and $X_2$, the number of ones for $X_1X_2$ K-code equals to the Hamming distance between $X_1$ and $X_2$. According to the proposed algorithm, the Hamming distance between $X_1$ and $X_2$ is either 3 or 4, thus $X_1X_2$ is placed in either $N_3$ or $N_4$. Based on our procedure so far, if $X_1$ and $X_2$ are both placed in $N_4$, the Hamming distance of $X_1$ and $X_2$ from $X_1X_2$ will be equal to 4. If $X_1$ and $X_2$ are both placed in $N_5$, the Hamming distance of $X_1$ and $X_2$ from $X_1X_2$ equals to 5. If either one of $X_1$ and $X_2$ is placed in $N_4$ and the other in $N_5$, then the Hamming distance of $X_1X_2$ from the one placed in $N_4$ will be 5 and 4 from the other one. Since $X_1X_2$ will not be placed on $X_1$ or $X_2$ and their First order side squares, it can be placed in the map without any interference. Now that $X_1$ and $X_2$ are placed in the map, we have to check different placements for $X_3$. To do so, we will first investigate the placement of $X_1X_3$ and $X_2X_3$. Since the least Hamming distance of $X_3$ from $X_1$ and $X_2$ is 3, $X_1X_3$ and $X_2X_3$ will be placed in $N_3$ or in $N_i$ where $i > 3$.

According to Theorem 3, $X_1X_3$ will not be placed on $X_1$ or $X_3$ or their First order side squares. Also, $X_2X_3$ will not be placed on $X_2$ or $X_3$ or their First order side squares. The only problem that might occur is in case of $X_2X_3$ being placed over $X_1$ or one of its First order side squares and in the same time $X_1X_3$ being placed over $X_2$ or one of its First order side squares. To avoid this situation, we know that if we have the K-code of $X_iX_j$ as well as the one for $X_i$ we can find the K-code of $X_j$. To find the K-code of $X_j$, we have to put the complement value of $X_iX_j$ where $X_i$ is none zero and put the same value of $X_iX_j$ in the K-code of $X_j$ where $X_i$ is 0. By using this procedure and assuming that the K-code of $X_1$ and each one of its First order side squares are equal to the K-code of $X_2X_3$ in separate cases, since the value of $X_2$ is known, the assumed value for $X_3$ can be found. To avoid putting $X_3$ in these squares we mark them for future reference. If we do the same procedure for each one of $X_2$ and its First order side squares with the assumption that they are equal to the K-code of $X_1X_3$ in separate cases, we will get to the same forbidden squares as before. The Karnaugh map of $S_{44}^4$ is represented in Fig. 2. The squares labeled as "$f$" are the forbidden squares for placement of $X_3$. Now, since $X_3$ should not be placed in forbidden squares (Side squares of $X_1$ and $X_2$ and $P_k$) and previously marked squares, we have to investigate different possibilities for placement of $X_3$. Since there are too many situations available for placement of $X_3$, we will only calculate the double weight side squares for one situation and just mention the value for the other placements. The first situation is $S_{444}^{444}$. In this situation, there will be 10 double weight side squares due to the placement of $X_3$ in $N_4$ which was already counted. In addition, 6 squares of Second order side squares of $X_3$, in accordance to Theorem 1, are placed over Second order side squares of $X_1$ from which, one placed in $N_2$, and 6 squares of Second order side squares of $X_3$ are placed over Second order side squares of $X_2$ from which, one is placed in $N_2$ and another one is the same as the common Second order side squares of $X_1$ and $X_3$. Therefore, the total number of double weight side squares of $X_3$ would be equal to 19. Since the procedure of calculating this number in different cases is familiar to each other we just provide it in Table. 1 for every possible case. Now, we have considered all possible cases for placement of $X_3$. In order to complete our algorithm, we will choose some of the situations for placement of $X_3$ in which the numbers of double weight side squares is larger. However, there are also other situations to consider that lead to reasonable results, but the probability of success in these situations is higher. Finally, our algorithm would be as follows: either all of $X_1$, $X_2$ and $X_3$ are placed in $N_4$ with a Hamming distance of 4, or one of them is placed in $N_5$ and other two are placed in $N_4$ in such a way that the one in $N_5$ has a Hamming distance of 3 from the other two, and the other two have a Hamming distance of 4 from each other. Another possible solution is to put, two of them in $N_5$ with a Hamming distance of 4 from each other and the other one in $N_4$ with a Hamming distance of 3 from the other two. Of course, none of them should be placed in any of forbidden squares. The Karnaugh map of $S_{445}^{433}$ is represented in Fig. 3. Now that the location for $X_1$, $X_2$ and $X_3$ are determined, we put $X_4$ in a place in such a way that neither $X_1$, $X_2$, $X_3$ or $P_k$ nor their side squares coincide with that square, and in a way that none of $X_4 X_l$ in which $l \in \{1,2,3\}$ are placed on $X_1$, $X_2$, $X_3$, $P_k$ or their side squares (The procedure for placement of $X_4$ is the same as what was done for $X_3$ considering $X_1 X_3$ and $X_2 X_3$). Note that, apart from this algorithm, any situation which results in a larger number of double weight side squares is of a higher priority.

| $s_6$ $s_4$ $s_7$ $s_2$ $s_5$ $s_3$ $s_1$ | 000 | 001 | 011 | 010 | 110 | 111 | 101 | 100 |
|---|---|---|---|---|---|---|---|---|
| 0000 | N | $P_2$ | $P_2P_4$ | $P_4$ | $P_4P_6$ | $X_1P_7$ | $P_2P_6$ | $P_6$ |
| 0001 | $P_1$ | $P_1P_2$ |  | $P_1P_4$ |  |  |  | $P_1P_6$ |
| 0011 | $P_1P_3$ |  |  |  |  |  |  |  |
| 0010 | $P_3$ | $P_2P_3$ |  | $P_3P_4$ | $f$ |  |  | $P_3P_6$ |
| 0110 | $P_3P_5$ | $X_2P_7$ |  | $f$ | $X_1X_2$ | $f$ |  | $f$ |
| 0111 |  |  |  |  | $f$ |  |  |  |
| 0101 | $P_1P_5$ |  |  |  |  |  |  |  |
| 0100 | $P_5$ | $P_2P_5$ |  | $P_4P_5$ | $f$ |  |  | $P_5P_6$ |
| 1100 | $P_5P_7$ | $X_2P_3$ |  |  |  | $X_1P_5$ |  |  |
| 1101 |  |  |  |  |  |  |  |  |
| 1111 |  | $X_2P_1$ |  |  |  |  |  |  |
| 1110 | $X_2P_2$ | **$X_2$** | $X_2P_4$ |  | $f$ |  | $X_2P_6$ |  |
| 1010 | $P_3P_7$ | $X_2P_5$ |  |  |  | $X_1P_3$ |  |  |
| 1011 |  |  |  |  |  |  |  |  |
| 1001 | $P_1P_7$ |  |  |  |  | $X_1P_1$ |  |  |
| 1000 | $P_7$ | $P_2P_7$ | $X_1P_6$ | $P_4P_7$ | $X_1P_2$ | **$X_1$** | $X_1P_4$ | $P_6P_7$ |

Fig. 2. Karnaugh map of $S_{44}^4$ and its forbidden squares for placement of $X_3$.

| Placement of data bits | The total number of double weight side squares of $X_3$ |
|---|---|
| $S_{445}^{455}$, $S_{454}^{365}$, $S_{544}^{356}$, $S_{554}^{455}$ | 10 |
| $S_{446}^{444}$, $S_{447}^{433}$, $S_{456}^{343}$, $S_{546}^{334}$ | 11 |
| $S_{444}^{446}$, $S_{444}^{464}$, $S_{445}^{435}$, $S_{445}^{453}$, $S_{454}^{345}$, $S_{544}^{354}$, $S_{454}^{363}$, $S_{544}^{336}$, $S_{455}^{354}$, $S_{545}^{345}$, $S_{554}^{435}$, $S_{554}^{453}$ | 15 |
| $S_{444}^{444}$, $S_{445}^{433}$, $S_{454}^{343}$, $S_{544}^{334}$, $S_{455}^{334}$, $S_{545}^{343}$, $S_{554}^{433}$ | 19 |
| Other cases | Invalid |

Table. 1. The total number of double weight side squares of $X_3$ for different cases.

## IV. INCREASING MAP CAPABILITIES

| $s_6$ $s_4$ $s_7$ $s_2$ $s_5$ $s_3$ $s_1$ | 000 | 001 | 011 | 010 | 110 | 111 | 101 | 100 |
|---|---|---|---|---|---|---|---|---|
| 0000 | N | $P_2$ | $P_2 P_4$ | $P_4$ | $P_4 P_6$ | $X_1 P_7$ | $P_2 P_6$ | $P_6$ |
| 0001 | $P_1$ | $P_1 P_2$ |  | $P_1 P_4$ |  |  |  | $P_1 P_6$ |
| 0011 | $P_1 P_3$ |  | $X_3 P_7$ |  |  |  |  | $X_1 X_3$ |
| 0010 | $P_3$ | $P_2 P_3$ |  | $P_3 P_4$ |  |  |  | $P_3 P_6$ |
| 0110 | $P_3 P_5$ | $X_2 P_7$ |  |  | $X_1 X_2$ |  |  |  |
| 0111 |  |  |  |  |  |  |  |  |
| 0101 | $P_1 P_5$ |  |  | $X_2 X_3$ |  |  |  |  |
| 0100 | $P_5$ | $P_2 P_5$ |  | $P_4 P_5$ |  |  |  | $P_5 P_6$ |
| 1100 | $P_5 P_7$ | $X_2 P_3$ |  |  |  | $X_1 P_5$ |  |  |
| 1101 |  |  |  |  |  |  |  |  |
| 1111 |  | $X_2 P_1$ | $X_3 P_5$ |  |  |  |  |  |
| 1110 | $X_2 P_2$ | **$X_2$** | $X_2 P_4$ |  |  |  | $X_2 P_6$ |  |
| 1010 | $P_3 P_7$ | $X_2 P_5$ | $X_3 P_1$ |  |  | $X_1 P_3$ |  |  |
| 1011 |  | $X_3 P_4$ | **$X_3$** | $X_3 P_2$ |  | $X_3 P_6$ |  |  |
| 1001 | $P_1 P_7$ |  | $X_3 P_3$ |  |  | $X_1 P_1$ |  |  |
| 1000 | $P_7$ | $P_2 P_7$ | $X_1 P_6$ | $P_4 P_7$ | $X_1 P_2$ | **$X_1$** | $X_1 P_4$ | $P_6 P_7$ |

Fig. 3. Karnaugh map of $S_{445}^{433}$.

In this section we will investigate correction of some three bit errors in transmission code. Note that we have to allocate an unassigned square of the map to a unique three bit error situation. Prior to open this discussion we will prove a theorem that will help us to provide our desired result.

*Theorem 4:* All three bit errors in form of $P_l P_m P_n$ in which $l \neq m \neq n$ and $l, m, n \in \{1, ..., 7\}$ cannot be put in a map containing all the possible situations of 2 bit errors in the ten bit code consisting of 3 data and 7 parity bits, in which each square would be occupied by only one error case.

*Proof:* By putting all $P_l P_m P_n$ in the map, all the $N_3$ squares will be filled. This makes it impossible for $X_1$, $X_2$ and $X_3$ to be placed in $N_4$, since the number of ones for $X_i X_j$ K-code is equal to the Hamming distance of $X_i$ and $X_j$, the Hamming distance between $X_1$, $X_2$ and $X_3$ has to be at least 4. Therefore, the only possible situation is that all three $X_1$, $X_2$ and $X_3$ be placed in $N_5$ and have a Hamming distance of 4 from each other. In this case, $X_1 X_2$ would have a Hamming distance of 1 from $X_3$, $X_1 X_3$ would have a Hamming distance of 1 from $X_2$ and also $X_2 X_3$ would have a Hamming distance of 1 from $X_1$ which is invalid. Therefore, all three bit errors in form of $P_lP_mP_n$ cannot be placed in this map. □

Now we have to note that, in order to cover all the three bit errors, in addition to the fact that the Hamming distance of $X_i$ and $X_j$ from each other has to be 5, they also have to be placed in $N_5$. In maps with 8, 9 or 10 parity bits, we can find situations to meet the mentioned conditions, but in these situations, $X_iX_j$ or $X_iX_jP_k$ are placed in invalid squares, therefore in order to cover all three bit errors we will need a map with at least 11 parity bits. This seems too large for the scope of this research. Now, we try to find a situation in a map with 7 parity bits which can cover the largest number of three bit errors. In $S_{444}^{444}$ the total number of three bit errors would be equal to 36 from which 9 cases are for $X_iX_jP_k(i,j\epsilon\{1,2,3\} \& k\epsilon\{1,...,7\})$, 11 for $P_lP_mP_n$, 15 for $X_iP_kP_m$ and one for $X_1X_2X_3$. In $S_{445}^{433}$, $S_{454}^{343}$ and $S_{544}^{334}$ the total number of three bit errors would be equal to 45 from which 11 cases are for $X_iX_jP_k$, 13 for $P_lP_mP_n$ and 21 for $X_iP_kP_m$. In $S_{455}^{334}$, $S_{545}^{343}$ and $S_{554}^{433}$ the total number of three bit errors would be equal to 38 from which 11 cases are for $X_iX_jP_k$, 8 for $P_lP_mP_n$ and 19 for $X_iP_kP_m$. In $S_{447}^{433}$, $S_{474}^{343}$ and $S_{744}^{334}$ the total number of three bit errors would be equal to 49 from which 4 cases are for $X_iX_jP_k$, 21 for $P_lP_mP_n$ and 24 for $X_iP_kP_m$. In $S_{456}^{343}$, $S_{645}^{433}$, $S_{564}^{334}$, $S_{465}^{433}$, $S_{546}^{334}$ and $S_{654}^{343}$ the total number of three bit errors would be equal to 40 from which 4 cases are for $X_iX_jP_k$, 17 for $P_lP_mP_n$ and 19 for $X_iP_kP_m$. In $S_{446}^{444}$, $S_{464}^{444}$ and $S_{644}^{444}$ the total number of three bit errors would be equal to 39 from which 5 cases are for $X_iX_jP_k$, 17 for $P_lP_mP_n$ and 17 for $X_iP_kP_m$. In $S_{556}^{433}$, $S_{565}^{343}$ and $S_{655}^{334}$ the total number of three bit errors would be equal to 38 from which 5 cases are for $X_iX_jP_k$, 11 for $P_lP_mP_n$ and 22 for $X_iP_kP_m$. In $S_{445}^{435}$, $S_{454}^{345}$, $S_{544}^{354}$, $S_{445}^{453}$, $S_{454}^{543}$ and $S_{544}^{534}$ the total number of three bit errors would be equal to 37 from which 7 cases are for $X_iX_jP_k$, 13 for $P_lP_mP_n$ and 17 for $X_iP_kP_m$. Any other situation is not possible and considered invalid. The map which contains the most three bit errors that would be one of $S_{447}^{433}$ or $S_{474}^{343}$ or $S_{744}^{334}$ is presented in Fig. 4.

Since three bit burst errors in sending code is of high importance and contains a high probability to occur, we will use the provided pattern in [2] to find similar patterns for placement of data bits and parity bits in sending code in which the map provided in Fig. 5 contains all the three bit burst errors. In order to achieve this purpose, if we transmit the sending code as in form of $X_1P_7P_3P_6X_3P_2P_4P_1P_5X_2$, we figure out that all the three bit burst errors ($X_1P_7 P_3, P_7P_3P_6, P_3P_6 X_3, P_6X_3 P_2, X_3P_2P_4, P_2P_4P_1, P_4P_1P_5$ and $P_1P_5X_2$) are available in the map, and we can simply reach the pattern of $X_1P_1P_2P_3X_2P_4P_5P_6P_7X_3$ for the sending code by displacing parity bits and data bits. In addition, if our code be in form of $X_1P_2P_5X_3P_4P_3P_1P_6P_7X_2$, again all the three bit burst errors will be available in the map and we can still get $X_1P_1P_2X_2P_3P_4P_5P_6P_7X_3$ by displacing parity bits and data bits. And finally, if the code is considered as $X_2P_5X_3P_2P_4P_3P_7P_6P_1X_1$, we realize that all three bit burst errors can be found in the map and we can achieve the pattern $X_1P_1X_2P_2P_3P_4P_5P_6P_7X_3$ for the sending code by changing data bits and parity bits. Within our study, we could not extract any pattern other than three above and the one provided in [2] from this map.

## V. CONCLUSION

In this paper, an algorithm was provided in order to place data bits in Karnaugh map by using a geometric approach and algebraic methods. The main objective for presenting this algorithm was to reduce the calculation time of computer in using Karnaugh map to detect and correct errors in sending code. The provided algorithm can be easily expanded for any desired amount of data bits.

Correction of three bit errors using Karnaugh map was investigated before. We have tried to cover a greater amount of three bit errors by expanding the map. We concluded covering all three bit errors using Karnaugh map is not affordable. Therefore, we provided a map which can cover the most three bit errors and used that map to create patterns to cover three bit burst errors in transmission code.

## REFERENCES

[1] R.K.Ward, M.Tabandeh. "Error Correction and detection a geometric approach" the computer journal, 27(3).pp.24 6-253 (1984)
[2] M.Tabandeh. "application of Karnaugh map for easy generation of correcting codes" *Scientia Iranica*, *Volume 19, Issue 3*, *June 2012*, *Pages 690-695.*
[3] C.L.Chen, "Error Correcting Codes with Byte Error-Detection Capability", IEEE Trans. On Computers, Vol.C-32, no.7, July, 1983.
[4] Reviriego, P., Maestro, J.A., O'Donnell, A. and Bleakley, C.J. "Soft error detection and correction for FFT based convolution using different block lengths", 15th IEEE International on-Line Testing Symposium, pp. 139-143 (June 2009).
[5] M. Schwartz, "Quasi-Cross Lattice Tilings with Applications to Flash Memory", arXiv: 1102.2035v1.
[6] R.H.Morelos-zaragoza, The Art of Error Correcting Coding , Johnwiley , west Sussex (2002)
[7] T.Klave, Codes for Error Detection , World scientific , Singapore (2007)
[8] Ganji, Mehdi, and Hamid Jafarkhani. "Novel Time Asynchronous NOMA schemes for Downlink Transmissions." *arXiv preprint arXiv:1808.08665* (2018).
[9] D.k.Pradham and J.J.Stiffer, Error-correcting codes and self-checking circuits. IEEE, COMPUTER, March (1980).

[10] M.Rudelson and R.Vershynin, Geometric approach to error-correcting codes and reconstruction of signals, Int Math Res Notices, 2005; 2005: 4019 - 4041.

[11] S. Lin, D.J. Costello Jr. "Error Control Coding", (2nd ed.)Prentice Hall, Englewood Cliffs (2004).

| $s_6$ $s_4$ $s_7$ $s_2$ $s_5$ $s_3$ $s_1$ | 000 | 001 | 011 | 010 | 110 | 111 | 101 | 100 |
|---|---|---|---|---|---|---|---|---|
| 0000 | N | $P_2$ | $P_2P_4$ | $P_4$ | $P_4P_6$ | $X_1P_7$ | $P_2P_6$ | $P_6$ |
| 0001 | $P_1$ | $P_1P_2$ | $P_1P_2P_4$ | $P_1P_4$ | $X_2X_3$ | | $P_1P_2P_6$ | $P_1P_6$ |
| 0011 | $P_1P_3$ | $P_1P_2P_3$ | | $P_1P_3P_4$ | $X_2X_3P_3$ | $X_3P_5P_7$ | | $P_1P_3P_6$ |
| 0010 | $P_3$ | $P_2P_3$ | $P_2P_3P_4$ | $P_3P_4$ | | $X_1P_3P_7$ | $P_2P_3P_6$ | $P_3P_6$ |
| 0110 | $P_3P_5$ | $X_2P_7$ | $X_2P_4P_7$ | | $X_1X_2$ | | $X_2P_6P_7$ | |
| 0111 | $X_1X_3$ | | $X_3P_6P_7$ | $X_1X_3P_4$ | | $X_3P_7$ | $X_3P_4P_7$ | $X_1X_3P_6$ |
| 0101 | $P_1P_5$ | $P_1P_2P_5$ | | $P_1P_4P_5$ | $X_2X_3P_5$ | $X_3P_3P_7$ | | $P_1P_5P_6$ |
| 0100 | $P_5$ | $P_2P_5$ | $P_2P_4P_5$ | $P_4P_5$ | | $X_1P_5P_7$ | $P_2P_5P_6$ | $P_5P_6$ |
| 1100 | $P_5P_7$ | $X_2P_3$ | | $P_4P_5P_7$ | $X_1P_2P_5$ | $X_1P_5$ | | $P_5P_6P_7$ |
| 1101 | $P_1P_5P_7$ | $X_2P_1P_3$ | $X_3P_3P_6$ | | $X_3P_2P_3$ | $X_3P_3$ | $X_3P_3P_4$ | |
| 1111 | | $X_2P_1$ | $X_3P_6$ | $X_3P_2P_6$ | $X_3P_2$ | **$X_3$** | $X_3P_4$ | $X_3P_2P_4$ |
| 1110 | $X_2P_2$ | **$X_2$** | $X_2P_4$ | $X_2P_2P_4$ | | $X_3P_1$ | $X_2P_6$ | $X_2P_2P_6$ |
| 1010 | $P_3P_7$ | $X_2P_5$ | | $P_3P_4P_7$ | $X_1P_2P_3$ | $X_1P_3$ | | $P_3P_6P_7$ |
| 1011 | $P_1P_3P_7$ | $X_2P_1P_5$ | $X_3P_5P_6$ | | $X_3P_2P_5$ | $X_3P_5$ | $X_3P_4P_5$ | |
| 1001 | $P_1P_7$ | $P_1P_2P_7$ | $X_1P_1P_6$ | $P_1P_4P_7$ | | $X_1P_1$ | $X_1P_1P_4$ | $P_1P_6P_7$ |
| 1000 | $P_7$ | $P_2P_7$ | $X_1P_6$ | $P_4P_7$ | $X_1P_2$ | **$X_1$** | $X_1P_4$ | $P_6P_7$ |

Fig. 4. Karnaugh map with the maximum three bit errors.

[12] D.Nikolos, "Theory and Design of t-Error Correcting/d-Error Detecting (d > t) and All Unidirectional Error Detecting Codes", IEEE Trans. On Computers, Vol.40 , no.2, February, 1991.

[13] Ganji, Mehdi, and Hamid Jafarkhani. "On the performance of MRC receiver with unknown timing mismatch-a large scale analysis." *arXiv preprint arXiv:1703.10422* (2017).

[14] Fooshee, David, Aaron Mood, Eugene Gutman, Mohammadamin Tavakoli, Gregor Urban, Frances Liu, Nancy Huynh, David Van Vranken, and Pierre Baldi. "Deep learning for chemical reaction prediction." *Molecular Systems Design & Engineering* (2018).

[15] Q. Huang, S. Lin, and K. A. S. Abdel-Ghaffar, "Error-correcting codes for flash coding, " *IEEE Trans. Inf. Theory*, vol. 57, no. 9, pp. 6097–6108, Apr. 2011.

[16] Shahosseini, S., Moazzemi, K., Rahmani, A. M., & Dutt, N. (2017, October). Dependability evaluation of SISO control-theoretic power managers for processor architectures. In *Nordic Circuits and Systems Conference (NORCAS): NORCHIP and International Symposium of System-on-Chip (SoC), 2017 IEEE* (pp. 1-6). IEEE.

[17] R. H. Morelos Zaragoza, *The Art of Error Correcting Coding*, 2nd Edition, John Wiley & Sons, 2006.

[18] E. Yaakobi, P. H. Siegel, A. Vardy, and J. K.Wolf, "On codes that correct asymmetric errors with graded magnitude distribution", in *Proc. IEEE Int. Symp. Inf. Theory*, St. Petersburg, Russia, Aug. 2011, pp. 1056–1060.

[19] J. Mathew, A. M. Jabir, H. Rahaman, D. K. Pradhan, "Single error Correctable Bit Parallel Multipliers over GF($2^m$) ", IET Comput. Digit. Tech., Vol. 3, Iss.3, pp. 281-288, 2009.

[20] O. Keren "One-to-Many: Context-Oriented Code for Concurrent Error Detection" Journal of Electron Test, vol. 26, pp. 337-353, 2010.

[21] Ganji, Mehdi, and Hamid Jafarkhani. "Interference mitigation using asynchronous transmission and sampling diversity." *arXiv preprint arXiv:1609.07032* (2016).

**Pouya Pezeshkpour** was born in Isfahan, Iran, in 1991. He is a B.S. under graduate student in electrical engineering department at Sharif University of Technology.

**Mahmoud Tabandeh** received his electronic engineering degree from Institut National des Sciences Appliquées (INSA) de Lyon (France), his M.S. degree from Louisiana State University (LSU) and his Ph.D. degree from the university of California, Berkeley (UCB).

He is currently an associate Professor in the School of Electrical Engineering, Sharif University of Technology, Tehran, Iran. His research interests include digital systems, hardware and software in general and image processing in particular.